\documentclass[
 aps, pra,
 amsmath,amssymb,
 11pt,
 final,
tightenlines,
 twoside,
 twocolumn,
 nofloats,
nofootinbib,
 superscriptaddress,
showkeys,
showkeywords,
 ]
{revtex4}

\usepackage[T2A]{fontenc}
\usepackage[utf8x]{inputenc}
\usepackage[russian,english]{babel}
\usepackage{graphicx}
\usepackage{dcolumn}
\usepackage{bm}

\usepackage{ulem}

\input{maik.rty}

\setcitestyle{authoryear,round}
\def\squareforqed{\hbox{\rlap{$\sqcap$}$\sqcup$}}

\def\sq{\ifmmode\squareforqed\else{\unskip\nobreak\hfil
\penalty50\hskip1em\null\nobreak\hfil\squareforqed
\parfillskip=0pt\finalhyphendemerits=0\endgraf}\fi}

\def\utw{\smash{\rlap{\lower5pt\hbox{$\sim$}}}}

\def\udtw{\smash{\rlap{\lower6pt\hbox{$\approx$}}}}

\def\diameter{{\ifmmode\mathchoice
{\ooalign{\hfil\hbox{$\displaystyle/$}\hfil\crcr
{\hbox{$\displaystyle\mathchar"20D$}}}}
{\ooalign{\hfil\hbox{$\textstyle/$}\hfil\crcr
{\hbox{$\textstyle\mathchar"20D$}}}}
{\ooalign{\hfil\hbox{$\scriptstyle/$}\hfil\crcr
{\hbox{$\scriptstyle\mathchar"20D$}}}}
{\ooalign{\hfil\hbox{$\scriptscriptstyle/$}\hfil\crcr
{\hbox{$\scriptscriptstyle\mathchar"20D$}}}}
\else{\ooalign{\hfil/\hfil\crcr\mathhexbox20D}}%
\fi}}

\newcommand{\ab}{Astrophysical Bulletin }

\newcommand{\aj}{Astron.~J. }
\renewcommand{\apj}{Astrophys.~J. }
\newcommand{\apjs}{Astrophys.~J. Suppl. }

\newcommand{\mnras}{Monthly Notices Royal Astron. Soc. }

\newcommand{\rmxaa}{Revista Mexicana Astronom. Astrof\'{\i}s. }

\renewcommand{\nat}{Nature }

\newcommand{\apjl}{Astrophys.~J.}

\def\be{\begin{equation}}
\def\ee{\end{equation}}
\def\ba{\begin{eqnarray}}
\def\ea{\end{eqnarray}}

\def\msun{M_\odot}

\def\ltsima{$\; \buildrel < \over \sim \;$}
\def\simlt{\lower.5ex\hbox{\ltsima}}
\def\gtsima{$\; \buildrel > \over \sim \;$}
\def\simgt{\lower.5ex\hbox{\gtsima}}
\def\etal{{et al.,}}

\usepackage[dvipsnames]{color}

\begin{document}

\selectlanguage{english}



\title{Velocity Dispersion and H$\alpha$ emission of Ionized Gas in Star-forming Regions}

\author{\firstname{E.~O.}~\surname{Vasiliev}}
 \email{eugstar@mail.ru}
 \affiliation{Lebedev Physical Institute, Russian Academy of Sciences, Moscow, 117997 Russia}

\author{\firstname{Yu.~A.}~\surname{Shchekinov}}
 \affiliation{Raman Research Institute, Sadashiva Nagar, Bangalore, 560080 India}

\begin{abstract}
For understanding the nature of gaseous flows in star-forming regions of nearby galaxies it is usually utilized the relation between surface brightness in H$\alpha$ line and velocity dispersion of ionized gas known as ''surface brightness -- velocity dispersion diagram''. Using the three-dimensional gasdynamic simulations we consider the evolution of the synthetic diagrams for supershells driven by multiple supernova explosions in stellar cluster located in the galactic disk. We investigate the shape and structure of the diagrams depending on gaseous density and metallicity, disk scale height values. We show that there are several loci of values typical for young bubbles evolved in dense or rarefied gas at large heights above the disk midplane. We find that the structure of the diagram is depended on age of a supershell and physical conditions of a gas in the disk. We argue that the diagrams obtained for the nearby dwarf galaxies can be interpreted by only using the dynamics of bubbles driven by multiple supernova explosions in low-mass stellar clusters of different age.
\end{abstract}

\maketitle

\section{INTRODUCTION}

Three-dimensional spectroscopy in optical emission lines makes it possible to study the kinematics of ionized interstellar gas in nearby galaxies with a spatial resolution of $5-50$~pc $1-3''$ \citep[for example,][]{munoz96,yang96,moiseev12}. Such observations reveal extended structures in the vicinity of star formation regions, where ionizing radiation from massive stars and multiple supernova (SN) explosions in star clusters lead to the appearance of turbulent gas flows and the formation of large-scale bubbles and outflows from galactic disks. The structures of this type are most clearly manifested in dwarf galaxies \citep[for example,][etc]{puche92,walter99,weisz09,bagetakos11,egorov17}. 

To understand the nature of gas flows in star-forming regions, it was proposed to use the ratio between the surface brightness values in the H$\alpha$ line and the velocity dispersion, defined as the standard deviation of the Gaussian profile of this line after taking into account instrumental effects and subtracting thermal broadening \citep{munoz96,yang96}. Subsequently, using three-dimensional
spectroscopy, ''surface brightness--velocity dispersion'' diagrams of ionized gas $I({\rm H\alpha})-\sigma$ were obtained for a significant number of nearby dwarf galaxies, in which many star formation regions were identified \citep[for example,][]{martinez07,borlado09,moiseev10,moiseev12,egorov14,egorov17,egorov21}. 

The analysis of $I({\rm H\alpha})-\sigma$ diagrams was limited by the fact that the interpretation of such diagrams is based on a qualitative understanding of their structure, proposed by \citet{munoz96} and generalized by \citet{moiseev12}. The study of \citet{vms15} was the first attempt to construct $I({\rm H\alpha})-\sigma$ synthetic diagrams based on numerical models of the dynamics of interacting SN remnants. In particular, it was shown that high gas velocity dispersions are associated with the collision of young SN remnants. The degrading of spatial (angular) resolution leads to some decrease of the velocity dispersion, both for observational data of the galaxy IC~10 \citep{moiseev12} and for numerical calculations of interacting SN remnants \citep{vms15}. This effect is important for understanding the velocity dispersion values measured for dwarf galaxies observed at low resolution, such as DDO~53 and DDO~190. A detailed study of the influence of angular resolution on the estimation of the expansion rate of the wind shell based on an analytical model indicates an underestimation of this value, which leads to an overestimation of the kinematic age and an underestimation of the energy inflow required for its formation \citep{smirnov21}. 

Large-scale shells with a size of about $\sim 0.1-1$~kpc \citep{bagetakos11}, which are usually associated with cumulative SN explosions in star clusters, are clearly distinguished from the observational intensity maps in the H$\alpha$ line. Therefore, it is important to study the structure of $I({\rm H\alpha})-\sigma$ synthetic diagrams for bubbles formed by multiple SN explosions in galactic disks, as well as to study the manifestations of evolutionary changes in the diagram for various conditions in interstellar gas. This work examines the possibility of interpreting the $I({\rm H\alpha})-\sigma$ diagrams using only the properties of gas flows arising from the expansion of the shell, which is supported by SN explosions in a star cluster. Section 2 describes the model and initial conditions. Section 3 demonstrates the results. Section 4 discusses the application of the results and their implications. Section 5 summarizes the main results.

\section{MODEL DESCRIPTION}

\begin{table*}
\caption{Models.}
\center
\begin{tabular}{ccccc}
\hline
\hline
Model   & $n_0$, cm$^{-3}$ & [Z/H] & $z_0$, kpc & Grid (x,y,z), kpc$^3$, Number of cells  \\
\hline
 {\bf M1}     &  0.1              &  -0.5 &  0.2       & $2.4\times 2.4\times 4.8$, $608\times 608\times 1216$  \\
 {\bf M2}     &  0.1              &  -0.5 &  0.5       & $1.5\times 1.5\times 4.1$, $384\times 384\times 1024$  \\
 {\bf M3}     &  0.9              &   0.0 &  0.5       & $0.93\times 0.93\times 1.5$, $232\times 232\times 1536$   \\
\hline
\hline
\end{tabular}%
\label{tab-models}
\end{table*}

We studied the emission characteristics of ionized gas in shells formed by multiple SN explosions in a star cluster located in the galactic disk using a numerical solution of three-dimensional gas dynamics equations taking into account radiative cooling in Cartesian geometry. The dynamics of these shells and the initial conditions are described in detail in the study of \citet{vs22}. Here we demonstrate the main parameters of our model.

We assumed that the gas disk is initially in hydrostatic equilibrium in the gravitational potential \citep[see, for example,][etc]{avillez00,hill12,walsch15,li17,vsn19}, which consists of two components: a dark matter (DM) halo and a baryonic disk. The DM halo profile is taken from \citet{nfw}, the virial radius is 30~kpc, the concentration parameter $c=4.5$, which is close to the values for the dwarf galaxy Holmberg~II \citep{puche92}. For a stellar disk, the acceleration is perpendicular to its plane and is $g_*(z) = 2\pi G \Sigma_* {\rm tanh} (z/z_*)$, where $\Sigma_*$ is the stellar surface density and $z_*$ is the stellar disk scale height. The contribution from the gas disk is added by dividing $g_*(z)$ by the factor $f_* = \Sigma_*/(\Sigma_* + \Sigma_{gas})$ \citep{li17}, i.e., it is implicitly assumed that the self-gravity of the gas disk is not taken into account. Note that this does not have any effect on the global dynamics of the bubble formed by SN explosions. The gas number density in the plane of the disk is $n_0$, the gas temperature is assumed to be the same in all models and equal to $9\times 10^3$~K. The metallicity of the gas [Z/H] in the calculations remains constant throughout the whole computational domain. The values of $n_0$ and [Z/H] are ranged in the intervals $0.1-1$~cm$^{-3}$ and --1 to 0, respectively. At asymptotically large heights above the plane of the disk ($z$ is much greater than the hydrostatic equilibrium scale), the gas density is assumed to be uniform and equal to $10^{-3}$~cm$^{-3}$. The stellar and gas surface densities and the stellar scale height are adopted to obtain following values of the gas disk scale height: $z_0 = 0.5$, 0.4, 0.3, 0.2~kpc at a fixed gas density in the plane of the disk. From more than twenty models, we choose three ones, which more clearly reflect the influence of the gas and disk properties. Their parameters are given in Table~\ref{tab-models}.

SNe in the cluster are distributed randomly, and the radius of the cluster is 30~pc. The random positions and times of SN explosions are calculated once, during the initialization so that in all models the same configuration is considered: SNe explode at the same times and in the same places. When a SN explodes, mass and energy are injected into a small region. Its size is 4~pc; for the fiducial spatial resolution of 4~pc, the region occupies one cell. The energy of one SN is $10^{51}$~erg (is added in the form of thermal energy). The masses of massive stars, the progenitors of SNe, are distributed randomly within the range of 8--40~$\msun$, according to the initial Salpeter mass function. The number of massive stars is assumed to be 100, which corresponds to the total mass of the cluster of $M_\ast \sim 1.5\times 10^4~\msun$ (assuming one SN per 150~$\msun$). We begin our run when the most massive SN explodes. The intervals between subsequent SN explosions correspond to the average lifetime of massive stars, the value of which is related to stellar mass \citep{iben-book}. Calculations are carried out up to 26~Myr, which is slightly longer than the longest lifetime of massive stars, that is, about 24~Myr for an 8~$\msun$ star.

To solve numerically the equations of gas dynamics, we use the explicit unsplit total variation diminishing (TVD) approach that provides high-resolution capturing of shocks and prevents unphysical oscillations. The scheme is of the Monotonic Upstream-Centered Scheme for Conservation Laws (MUSCL--Hancock) type. To improve accuracy when calculating flows at cell boundaries, the approximate Harten-Lax-van Leer Contact (HLLC) method is used to solve the Riemann problem \citep[e.g.,][]{toro99}. The code successfully passed the whole set of gas dynamic tests proposed in the paper of \citet{klingenberg07}.

To account for radiation losses in the calculations, we use a nonequilibrium cooling function \citep{v11,v13}, obtained for the isochoric process of gas cooling from $10^8$~K to 10~K, including the ionization kinetics of all ionic states of the following chemical
elements: H, He, C, N, O, Ne, Mg, Si and Fe. The gas can be heated by photoelectric heating of dust particles \citep{dust-heat}; this process is considered to dominate the heating of interstellar gas. Any deviation in the heating rate in unperturbed gas violates the balance between cooling and heating, stimulates the development of thermal instability, and leads to redistribution of the mass of interstellar gas in the disk \citep[for example,][]{avillez00,hill12}. To avoid the influence of these effects in the calculations, we assume an exponential decrease in the heating rate in the direction perpendicular to the plane of the disk with a characteristic scale equal to the gas scale height of the disk \citep{li17}. Such assumptions make it possible to successfully stabilize the radiative cooling of the surrounding gas at $T=9\times 10^3$~K at the initial moment and obtain an equilibrium gas disk on a time scale exceeding the calculation time.

To obtain a data cube in the H$\alpha$ line, the gas emissivity of each element (grid cell) determined by the gas temperature and H II fraction is calculated. The value of the latter at a given temperature is found from the precomputed tables for the corresponding cooling functions \citep{v13}. Next, along each line of sight, the emission in the H$\alpha$ line is integrated in the velocity range $(v,v+\Delta v)$ with a resolution of $\Delta v=1$~km~s$^{-1}$. The total intensity is found by summing over all velocity intervals.

\section{RESULTS}

The evolution of a bubble formed by multiple SN explosions in a star cluster located in the plane of the disk has been described in many works \citep[for example,][and many others]{castor75,maclow88,vsn17,fielding18}, so we will not go into detail on this. However, it should be noted that in our previous work \citep{vs22} the dynamics of a bubble around a star cluster with a mass $M_* \sim 1.5\times 10^4\msun$ was studied depending on the value of the disc scale height. In particular, it was noticed that in a disk with a scale height $z_0 \simgt 0.5$~kpc, the resulting bubble remains almost spherical throughout the whole evolutionary time under consideration, that is, up to 26~Myr. For a smaller scale height, upon reaching a size comparable to the disk scale height,
the bubble begins to expand predominantly in the vertical direction. Thus, for $z_0=0.2$~kpc, the ratio of the maximum bubble size in the plane of the disk and in the plane perpendicular to it exceeds a factor of 1.5 by the time of 24~Myr \citep[see Fig.~1][]{vs22}. Despite of the larger size, the mass of the swept up gas in the vertical direction appears to be less: the shell high above the plane of the disk appears to be much thinner than in the plane of the disk. It should be pointed out that in these calculations the gas density in the plane of the disk was equal to 0.9~cm$^{-3}$, and the metallicity was assumed to be solar. As these values decrease, the
bubble reaches a size comparable to the scale height faster, and therefore the expansion of the bubble in the vertical direction will be more noticeable.

\begin{figure}
\center
\includegraphics[width=6cm]{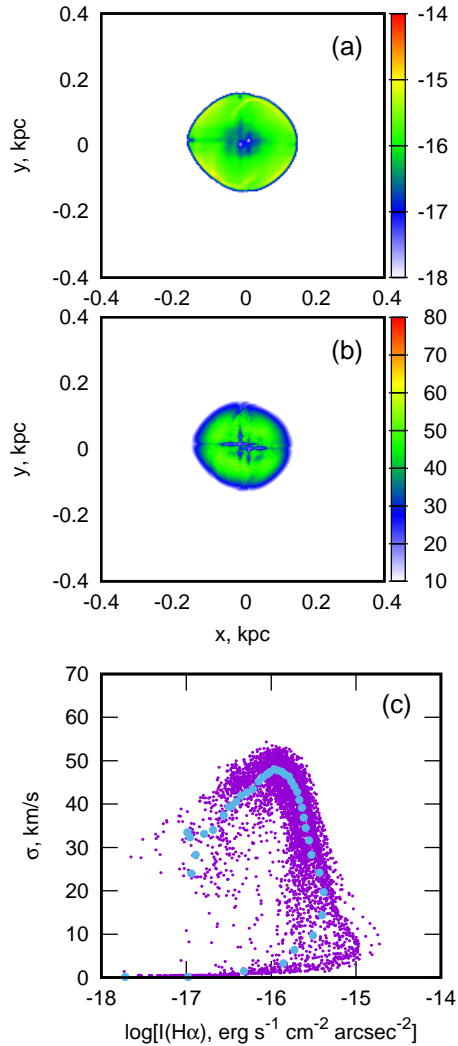}
\caption{
Surface brightness (a) and velocity dispersion (b) in the H$\alpha$ line of ionized gas in the supershell for {\bf M1} model: cluster in the plane of the disk with $z_0=0.2$~kpc and $n_0 = 0.1$~cm$^{-3}$, ${\rm [Z/H]}=-0.5$. The age is 1~Myr. The color scale in panel (a) is the logarithm of surface brightness in units of erg cm$^{-2}$ s$^{-1}$ arcsec$^{-2}$; in panel (b) the velocity is in km~s$^{-1}$. Panel (c) shows the ''surface brightness--velocity dispersion'' diagram of the ionized gas $I({\rm H\alpha})-\sigma$  of the distributions presented in panels (a) and (b), the blue dots correspond to the diagram for the values, averaged over annular regions with a center coinciding with the center of the cluster. 
}
\label{fig-maps}
\end{figure}

Using synthetic observations perpendicular to the plane of the disk we study the emission and dynamic characteristics of ionized gas in a bubble formed by multiple SN explosions in a star cluster with a mass of $M_* \sim 1.5\times 10^4\msun$. For example, Figure~\ref{fig-maps} shows\footnote{The intensity from the unperturbed part of the disk presents a uniform background with rather low intensity values presented by the white part of the color bar in panel (a). Vertical velocities in the unperturbed disk are almost absent, their dispersion is close to zero (see panel (b)).} the surface brightness (a) and velocity dispersion (b) in the H$\alpha$ line of ionized gas in a supershell formed by SN explosions in a cluster located in the plane of the disk with scale height $z_0=0.2$~kpc, midplane density $n_0 = 0.1$~cm$^{-3}$, metallicity ${\rm [Z/H]}=-0.5$: M1 model (see Table~\ref{tab-models}). The age of the supershell is 1~Myr. It can be seen that the surface brightness increases towards the border of the bubble, since the radial number density of gas is maximum there due to the formation of dense and thick walls the parts of the shell located above one scale height. At this time the average surface density of gas in the swept-off shell on the periphery is several times higher compared to the central regions. Later, the peripheral parts of the shell will become thicker and the central parts thinner, so the difference in column density and, therefore, in surface brightness will increase.

The velocity dispersion of the ionized gas towards the bubble border decreases, since the vertical component of the velocity decreases with radius\footnote{The cross-shaped structures in the central part are the result of the well-known numerical instability from recent SN explosions \citep{quirk94}; at this time their relative contribution to the total surface is several
percent, afterwards it rapidly decreases.}. In the central regions, the value of $\sigma$ appears to be $1.5-2$ times higher than in the peripheral regions, which are brighter in the H$\alpha$ line. The maximum velocity dispersion of about $\sim 50$~km~s$^{-1}$ corresponds to surface brightness $I({\rm H}\alpha) \sim 10^{-16}$~erg cm$^{-2}$ s$^{-1}$ arcsec$^{-2}$.

For clarity, Fig.~\ref{fig-maps}c shows the ''surface brightness–velocity dispersion'' diagram of the ionized gas ($I({\rm H\alpha})-\sigma$) for the distributions of values presented in panels (a) and (b). There are four regions in the diagram: i) the brightest regions with a velocity dispersion of about 10~km~s$^{-1}$ correspond to the bright border of the supershell; ii) a descending branch with velocity dispersion close to zero and brightness $\simlt 3\times 10^{-16}$~erg cm$^{-2}$ s$^{-1}$ arcsec$^{-2}$ corresponds to the outer edge of the supershell; iii) the ascending branch with a dispersion from 20 to 50~km~s$^{-1}$ and high brightness $I({\rm H}\alpha) \sim (0.3-3)\times 10^{-16}$~erg cm$^{-2}$ s$^{-1}$ arcsec$^{-2}$ represents the internal (excluding the central) parts of the supershell; iv) the lowest bright points with a wide spread of velocity dispersion are the central regions of the supershell (see panels (a) and (b) of Fig.~\ref{fig-maps}).

While the bubble expands, the $I({\rm H\alpha})-\sigma$ diagram changes significantly. Due to the spherical symmetry of supershells, it is possible to average the values of brightness and dispersion over an angle and to obtain radial profiles of the averaged values. To do this, we built a system of concentric rings around the center of the cluster. The width of each ring is 4~pc.

Figure~\ref{fig-maps}c demonstrates the dependence for averaged values (large blue symbols). The averaged emission values generally reflect quite well the features of the diagram for the full set of points. This allows us to consider the evolution of the diagram for angle-averaged surface brightness and velocity dispersion.

\citet{vms15} investigated the influence of spatial resolution degrading on diagram structure. In particular, both decrease in brightness and drop in velocity dispersion have beem found for lower resolution. This is explained by the fact that the sizes of areas with high brightness and dispersion values turn out to be small, on the order of $10-30$~pc, which is comparable to the averaging scale of $2.5-5$~pc. In the models studied here, the spatial variations in brightness and dispersion are greater, because supershells formed by multiple SN explosions are considered contrary to the interaction of individual SN remnants with each other. Therefore, we can expect that a decrease in resolution changes the structure of the diagram slightly.

\begin{figure*}
\center
\includegraphics[width=6.5cm]{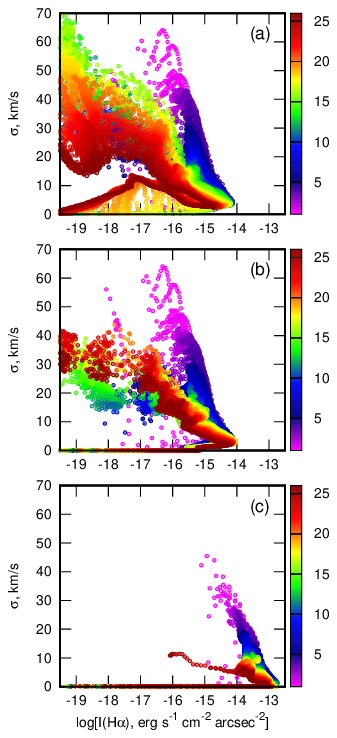}
\includegraphics[width=5.75cm]{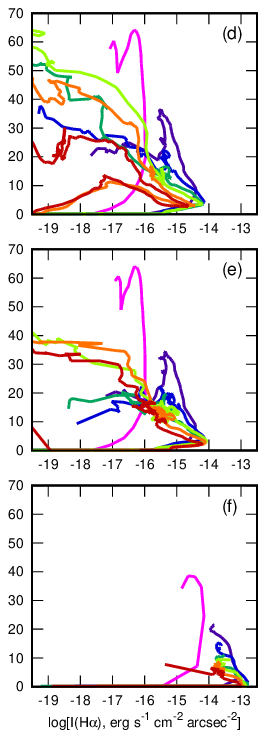}
\caption{
The ''surface brightness--velocity dispersion'' diagrams of ionized gas ($I({\rm H\alpha})-\sigma$) for spatial distributions averaged over annular regions with centers coinciding with the center of the cluster in the plane disk. The panels (a), (c) and (e) correspond to the models {\bf M2}, {\bf M2} and {\bf M3}, respectively. The color scale corresponds to the age of the supershell in Myr. The panels (b), (d), (f) show the same diagrams as in the panels (a), (c) and (e), but for time moments 0.4, 4, 8, 12, 16, 20 and 24~Myr. The line colors correspond to the color bar in the panels (a), (c), (e).
}
\label{fig-eha-sigma-rings}
\end{figure*}

Figure~\ref{fig-eha-sigma-rings} contains diagrams for the values averaged over concentric regions around a star cluster located in the plane of the disk. Let us consider the evolution of the diagram for a bubble in the {\bf M1} model, i.e., in a disk with scale height $z_0=0.2$~kpc and central density $n_0=0.1$~cm$^{-3}$, metallicity ${\rm [Z/H]}=-0.5$ (panel (a)). As it is noted above, by the time $t\sim 1$~Myr the bubble reaches a radius of the order of one scale height $r\sim z_0 \sim 0.2$~kpc (see Fig.~\ref{fig-maps}), the velocity dispersion in the inner parts of the supershell reaches 70~km~s$^{-1}$, the average brightness of these regions is $I({\rm H}\alpha) \sim 10^{-16}$~erg cm$^{-2}$ s$^{-1}$ arcsec$^{-2}$. During next few million years the velocity dispersion gradually decreases to $20-30$~km~s$^{-1}$.

\begin{figure}
\center
\includegraphics[width=6.5cm]{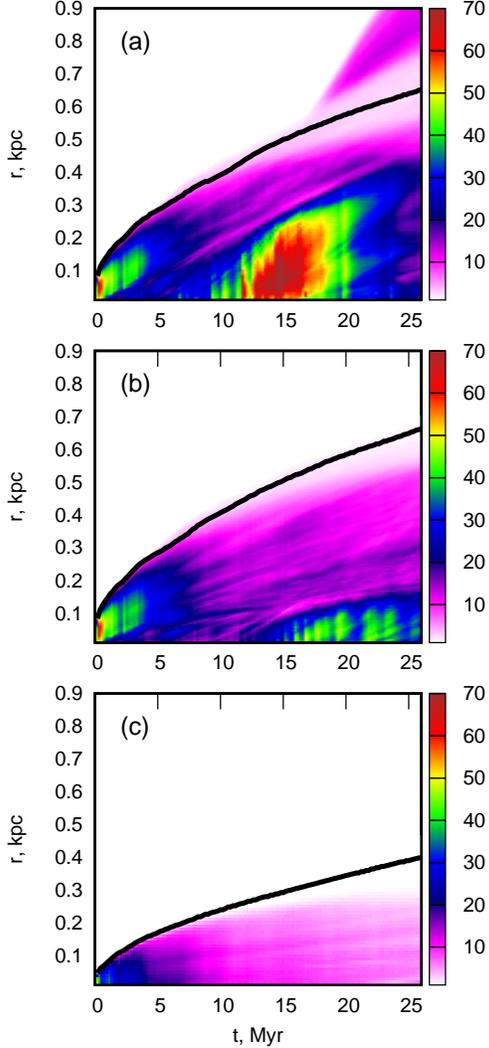}
\caption{
Space-time distribution of the velocity dispersion of the ionized gas $\sigma$ in the supershell (color scale in km~s$^{-1}$), averaged over concentric regions with a center coinciding with the center of the cluster, for parameters as in Fig.~\ref{fig-eha-sigma-rings}. The black solid line corresponds to the evolution of the supershell radius in the plane of the disk.
}
\label{fig-sigma-rings}
\end{figure}

By $t\sim 10$~Myr, the size of the bubble above the disk plane becomes larger than two scale heights of the disk. As a result, in further epoch ($t\simgt 11$~Myr), the expansion becomes accelerated in the vertical direction and the velocity dispersion increases. In the diagram this corresponds to the regions with dispersion of $30-40$~km~s$^{-1}$ and brightness of $\sim 10^{-17}-10^{-16}$~erg cm$^{-2}$ s$^{-1}$ arcsec$^{-2}$, as well as $\sigma \sim 60-70$~km~s$^{-1}$ and rather low brightness of $\sim 10^{-19}$~erg cm$^{-2}$ s$^{-1}$ arcsec$^{-2}$. Based on the space-time distribution of velocity dispersion in the supershell since $t\sim 11$~Myr (Fig.~\ref{fig-sigma-rings}a), the high velocity dispersion values appear in the central region of the supershell, namely within $r\simlt 0.1$~kpc. Further, at distances up to the supershell radius $r\sim 0.3-0.5$~kpc, the emission and dynamic properties of the supershell are determined by the dense and cold parts: the velocity dispersion quickly drops to $\sim 5-20$~km~s$^{-1}$, the surface brightness is $I({\rm H}\alpha) \sim (0.2-1) 10^{-14}$~erg cm$^{-2}$ s$^{-1}$ arcsec$^{-2}$.

By $t\sim 15$~Myr, the rate of SN explosions decreases to a level not allowing to support the accelerated expansion of the supershell. Therefore, during next few million years, the velocity dispersion in the central region gradually decreases to $\sim 20-30$~km s${-1}$. The size of the region with higher dispersion grows while SN explosions occur, and till $t\sim 24$~Myr it reaches almost 0.4~kpc. In Fig.~\ref{fig-eha-sigma-rings} this region corresponds to low surface brightness $I({\rm H}\alpha) \simlt 3\times 10^{-17}$~erg cm$^{-2}$ s$^{-1}$ arcsec$^{-2}$. After the explosions cease, the dispersion drops below 20~km~s$^{-1}$ for the whole range of brightness values. It should be noted that the part of the bubble at large heights above the disk ($z\simgt 3z_0$) extends over distances larger than the radius of the bubble in the plane of the disk. In the model considered here this occurs after $t\sim 15$~Myr, thus, in the dispersion distribution (Fig.~\ref{fig-sigma-rings}a) there is a region with a low dispersion value ($\sigma \sim 10$~km~s$^{-1}$) at distances exceeding the size of the supershell.

Increasing disk scale height to $z_0 = 0.5$~kpc the evolution of the bubble during the first 6--7~Myr remains virtually unchanged. Therefore, the structure of the $I({\rm H\alpha})-\sigma$ diagram also does not contain noticeable differences compared to the distribution for a disk with scale height $z_0 = 0.2$~kpc in the {\bf M2} model (see Fig.~\ref{fig-eha-sigma-rings}c). Subsequently, the vertical size of the bubble does not reach values above $z_0$, so no noticeable outflow is formed and the velocity dispersion is around $\sim 10$~km s$^{-1}$. Only in a small central part of the supershell ($r\simlt 0.1$~kpc) it reaches a maximum of
$\sim 40$~km~s$^{-1}$ (see Fig.~\ref{fig-sigma-rings}b). In the diagram, this central part corresponds to regions with low intensity in the H$\alpha$ $I({\rm H}\alpha) \simlt 3\times 10^{-17}$~erg cm$^{-2}$ s$^{-1}$ arcsec$^{-2}$. 

An increase of gas density and/or metallicity leads to earlier transition to the radiative phase and deceleration of the bubble expansion, which reveals in higher emission in recombination lines and lower velocity dispersion. In a disk with large scale height, the shape of the expanding bubble remains close to spherical within its evolution. Under these conditions the supershell velocity dispersion monotonically decreases with time, that can be seen on the bubble evolving in a disk with scale height $z_0=0.5$~kpc, midplane density $n_0=0.9$~cm$^{-3}$ and metallicity ${\rm [Z/H]}=0$ in {\bf M3} model (see Fig.~\ref{fig-eha-sigma-rings}e). Due to the higher density the H$\alpha$ intensity increases more than an order of magnitude compared to the values in the models presented in Fig.~\ref{fig-eha-sigma-rings}a,c. The velocity dispersion decreases monotonically with time and radius from $\sim 40$~km~s$^{-1}$ to less than 5~km~s$^{-1}$ (see Fig.~\ref{fig-sigma-rings}c).

\section{DISCUSSION}

\begin{figure*}
\center
\includegraphics[width=12.5cm]{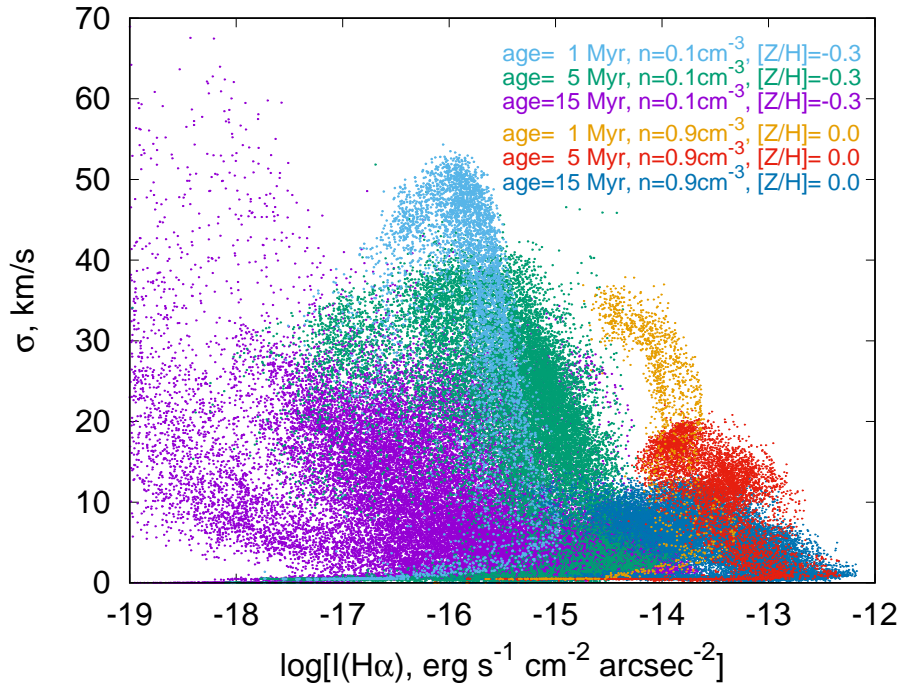}
\caption{
Composite diagram ''surface brightness--velocity dispersion'' of ionized gas ($I({\rm H\alpha})-\sigma$) for several supershells with ages of 1, 5 and 15~Myr, evolving in disks with $z_0=0.2$~kpc, $n_0=0.1$~cm$^{-3}$, ${\rm [Z/H]}=-0.5$  and $z_0=0.5$~kpc, $n_0=0.9$~cm$^{-3}$, ${\rm [Z/H]}=0$ (the color of the points corresponds to the different supershells, see the legend in the panel).
}
\label{fig-eha-sigma-scheme}
\end{figure*}

The ionization of interstellar gas in galaxies is the result of stellar radiation and shock waves. Therefore, at the ''surface brightness--velocity dispersion'' diagrams obtained from the observed spatial distributions, it is possible to identify regions corresponding to ionization sources of different natures \citep[for example,][]{munoz96,yang96,moiseev12}. This can be supported by a wide range of H$\alpha$ brightness values and velocity dispersion at the diagrams obtained from the observations of nearby dwarf galaxies \citep{moiseev12}. At the same time, one can be seen at the averaged synthetic diagrams in Fig.~\ref{fig-eha-sigma-rings} that the locus of points corresponded only to SNE shock waves at the diagrams appears to be close to observational ones. Diagrams for the superposition of SNe bubbles of different ages allow us to interprete the observations reasonably.

To carry out a simple analysis, we construct a composite diagram $I({\rm H\alpha})-\sigma$ for several supershells of different ages (Fig.~\ref{fig-eha-sigma-scheme}). This is possible due to stellar clusters and supershells of different ages are observed in the disks of dwarf galaxies \citep[see][for the Holmberg II galaxy]{egorov17}. Both disk scale height and midplane density vary with galactocentric radius \citep[for example,][]{bagetakos11}. Metallicity variations are also possible, since metals do not have time to mix during hundreds of Myr even in galaxies with high star formation rates (\citet[][]{avillez02} and discussions in \citet{andrews01,nasoudi10,decia21}).

First of all, it should be noted that the overall shape of this synthetic diagram is similar to some diagrams obtained from observations of dwarf galaxies \citep{moiseev12}: a quasi-triangular structure with apex at high H$\alpha$ brightness values around $10^{-12}$~erg cm$^{-2}$ s$^{-1}$ arcsec$^{-2}$ and low velocity dispersion values of more than 5~km~s$^{-1}$, a peak in velocity dispersion reaches 50~km~s$^{-1}$ at average values of $I({\rm H\alpha}) \sim 10^{-16}$~erg cm$^{-2}$ s$^{-1}$ arcsec$^{-2}$. The region with lower brightness characterized by a large scatter of velocity dispersion is not represented in the observations due to instrumental limitations.

More specifically,  supershells for given characteristics cane be predominantly connected with some regions of the composite diagram. The peaks of velocity dispersion about 50~km~s$^{-1}$ (blue dots) and 35~km~s$^{-1}$ (yellow dots) correspond to young supershells of age about 1~Myr, which are probably in the adiabatic phase or have just transited to radiative cooling. For example, the cooling time of a gas with density $n_0=0.1$~cm$^{-3}$ and metallicity ${\rm [Z/H]}=-0.5$ is $t_c \sim kT/4n_0\Lambda(T) \sim 1$~Myr for the gas temperature behind the shock wave front $T\sim 10^6$К. Thus, one can be found that the velocity dispersion is lower in gas with higher density and metallicity (compare blue and yellow points in the diagram). The sizes of these supershells are of the order of $50-150$~pc (see Fig.~\ref{fig-sigma-rings}). Similar velocity dispersion peaks can be found in the $I({\rm H\alpha})-\sigma$ diagrams for the galaxies DDO~53, DDO~99, VII~Zw~403, IC~10 and others, obtained using the Scorpio scanning Fabry-Perot interferometer on the BTA 6-m telescope \citep{moiseev12}. For these galaxies the velocity dispersion peaks correspond to H$\alpha$ intensities in the range from $10^{-17}$ to $3\times 10^{-15}$~erg cm$^{-2}$ s$^{-1}$ arcsec$^{-2}$. On H$\alpha$ maps the regions with higher velocity dispersion have sizes of several tens of parsecs \citep[see Fig. 1 and 4 in][]{moiseev12}, which is quite consistent with the simulated supershells.

For a supershell of 1~Myr old expanding in gas with $n_0=0.1$~cm$^{-3}$ and ${\rm [Z/H]}=-0.5$, the velocity dispersion varies from 0 to 50~km~s$^{-1}$ (blue dots in Fig.~\ref{fig-eha-sigma-scheme}): lower values correspond to the outer part of the supershell with a thickness of $10-20$~pc (see Fig.~\ref{fig-maps}b), which is comparable to the spatial resolution when observing nearby dwarf galaxies. During the bubble evolution the maximum velocity dispersion in the supershell decreases and reaches 40~km~s$^{-1}$, when it is expanding in a medium with $n_0=0.1$~cm$^{-3}$ and ${\rm [Z/H]}=-0.5$ (green dots) and 20~km~s$^{-1}$ for $n_0=0.9$~см$^{-3}$ и ${\rm [Z/H]}=0$ (red dots). Thus, the observed regions with high dispersion of ionized gas velocities $\sigma \simgt 30$~km~s$^{-1}$ can be associated with bubbles of about 1~Myr old formed due to SN explosions in stellar clusters $M_\ast \simgt 10^4\msun$ and expanding in gas with $n_0 \sim 0.1$~cm$^{-3}$ and ${\rm [Z/H]} \sim -0.5$. The sizes of these regions turn out to be quite compact (about $100-150$~pc). It should be noted that the estimates of the sizes obtained from observations probably do not include the region in the vicinity of the thick bubble shell, in which the velocity dispersion falls below 30 km s$^{-1}$ and whose thickness is $10-15$~pc, that corresponds to about 10\% of the bubble radius. Decrease of gas density and growth of the mass of the cluster reveal in increase of velocity dispersion. The thickness of the disk at such early phases of evolution does not play any role.

During 10 Myr since the explosion of the most massive star in the cluster, more than half of the SN have explodes (for the Salpeter initial mass function), so the SN rate can no longer support the expansion of the bubble at a high speed and the velocity
dispersion decreases to 10~km~s$^{-1}$. The H$\alpha$ brightness of such gas can vary over a fairly wide range; in particular, for a bubble in a medium with $n_0=0.9$~cm$^{-3}$ and ${\rm [Z/H]}=0$, the value of $I({\rm H\alpha})$ varies from $10^{-15}$ to $10^{-12}$~erg cm$^{-2}$ s$^{-1}$ arcsec$^{-2}$ (dark blue dots in Fig.~\ref{fig-eha-sigma-scheme}). It should be kept in mind that in this model the bubble expands in a disk with scale height $z_0=0.2$~kpc. Under such conditions the bubble radius reaches approximately $1.5z_0$ (Fig.~\ref{fig-sigma-rings}c) by 15~Myr. In a gas with lower density and/or metallicity, the radius of the bubble increases significantly by this time, for example, for $n_0=0.1$~cm$^{-3}$ and ${\rm [Z/H]}=-0.5$, it increases to
almost $2.5z_0$, which leads to a breakthrough of the disk. Under these conditions, this part of the shell expands rapidly, and its velocity dispersion increases. Unfortunately, the column density of gas in this part of the shell is low. At the composite diagram the locus of points corresponded to such gas is depicted by purple dots for high velocity dispersion.

Most of the purple points correspond to gas with low velocity dispersion of about $\sim 10-20$~km~s$^{-1}$, i.e. the gas located in the walls of the bubble expanding predominantly in the direction parallel to the disk plane at various heights above the plane (see regions marked by purple color at $t\simgt 15$~Myr in the top panel of Fig.~\ref{fig-sigma-rings}). After 10~Myr since SNe begin to explode, there is no gas with high velocity dispersion in the thick disk (see Fig. ~\ref{fig-sigma-rings}b). Therefore, the observed regions of high brightness and low velocity dispersion are likely to be associated with old bubbles.

The regions with low intensity in a wide range of velocity dispersion values, which appear in a breakthrough of the disk, are  associated with old bubbles as well. In general, for $\sigma$ below $\sim 10-15$~km~s$^{-1}$ it is quite difficult to determine the age of the supershell, since the contributions from bubbles of different ages are mixed in this region. However, for a pronounced region of high dispersion (a peak), the points with low dispersion and similar brightness, which are located below such region at the diagram, are evolutionarily and spatially related to each other: the region with high dispersion is enclosed by the region with low dispersion. Note that similar picture can be found in the galaxies DDO~53 for shells 1, 2 and 3 \citep[see Fig. 1 in][]{moiseev12} and  NGC~1385 \citep[see Fig. A1 in][]{egorov23}.

Once again, it is worth pointing out that low-brightness regions of the diagram are associated with a low-density turbulent interstellar medium, including regions of both low and high velocity dispersion \citep[see Fig. 6 in][]{moiseev12}. From the above results of numerical calculations one can conclude that supershells of various ages located in this region, for example, young remnants evolving in gas with low density, parts of old shells located both high above the plane of the disk, when it breaks through, and at low heights, expanding predominantly in a direction parallel to the disk plane. So that, this part of the diagram requires a more detailed separation of the components.

Finally, it should be noted that when calculating emission and dynamic features of the gas, it is assumed that the disk is seen face-on. Although, dwarf galaxies are at different angles to the observer. It is clear that for a edge-on galaxy one can observe a  different picture; the diagrams for such galaxies contain information about cumulative structure of the disk along the line of sight and vertical outflows from the disk (winds) \citep[see ][]{coba17}. In galaxies with inclination of $i \sim 27^\circ$, similar to  Holmberg~II \citep{sanchez14}, large-scale bubbles with quasi-sherical shapes in the HI 21 cm line are clearly seen. In the numerical models, the shape of the bubbles depends on disk scale height and evolutionary phase. It has already mentioned above that in thick disks the shape remains close to spherical throughout the whole evolution considered here. For thinner disks with low metallicity, after 12--15~Myr the shape of the shell becomes close to dumbbell-like, elongated in the vertical direction. In the models presented here, the opening angle of the vertical outflow is greater than $55^\circ$ (from the perpendicular to the plane of the disk). Thus,  our conclusions remain valid for smaller inclination.

The inhomogeneity of the medium can significantly influence on the dynamics of a bubble formed by large number of SN explosions only within one scale height. However, the shell velocity is tens of km~s$^{-1}$ at this scale, that leads to the destruction of the cloud structure. Only for highly mass-loading flows, in particular, when the density dispersion for the lognormal distribution is higher than $2\sigma$ (almost two orders of magnitude in density) the differences in the global dynamics of the bubble within the disk become remarkable. Such density fluctuations are rare in dwarf galaxies.

\section{CONCLUSIONS}

This paper examines the evolution of the ''surface brightness--velocity dispersion'' diagrams of ionized gas ($I({\rm H\alpha})-\sigma$) for supershells formed by multiple SNe explosions in a stellar cluster $M_\ast \sim 10^4\msun$, located in the midplane of the disk with scale height $z_0\sim0.2-0.4$~kpc. A variation of the shape and structure of the diagram has been studied depending on gas density and metallicity, disk scale height as well.

It has been shown that the structure of the $I({\rm H\alpha})-\sigma$ diagrams obtained from observations of nearby dwarf galaxies can be interpreted as follows:
\begin{itemize}
 \item the regions with high velocity dispersion of ionized gas $\sigma \simgt 30$~km~s$^{-1}$ can be associated with bubbles about 1~Myr old formed by SN explosions in stellar clusters $M_\ast \simgt 10^4\msun$ in gas with $n_0 \sim 0.1$~cm$^{-3}$ and ${\rm [Z/H]} \sim -0.5$; the sizes of these regions are quite compact, about $100-150$~pc; both decrease of gas density and increase of mass of a cluster lead to higher velocity dispersion; the thickness of the disk at such early phases of evolution does not play any role;
 \item in the region with velocity dispersion below 10-15~km~s$^{-1}$ it is quite difficult to determine the age of the supershell: the contributions from bubbles of different ages are mixed in it; in the presence of a pronounced region with high dispersion (peak) the points with lower dispersion and close brightness values, located under this region in the diagram, are evolutionarily and spatially related to each other: the region with high dispersion is enclosed by that with low dispersion;
 \item when the breakthrough of a disk occurs, the part of the shell expands rapidly and its velocity dispersion increases, however, gas column density and emission measure of this part of the shell are low; in the diagram this corresponds to the region with high velocity dispersion of several tens of km~s$^{-1}$ and low surface brightness of $I({\rm H\alpha}) \simlt 3\times 10^{-17}$~erg cm$^{-2}$ s$^{-1}$ arcsec$^{-2}$;
 \item the regions with high brightness and velocity dispersion below 10~km~s$^{-1}$ correspond to the shells older than 15~Myr evolving in denser gas, $n_0 \sim 1$~cm$^{-3}$. 
\end{itemize}

Thus, the structure of the $I({\rm H\alpha})-\sigma$ diagram for nearby dwarf galaxies can be completely explained by using solely the dynamics of bubbles formed by multiple SN explosions in small stellar clusters of different ages. The analysis of the emission properties of collective SNe shells from OB associations provides a resonable model for interpreting the $I({\rm H\alpha})-\sigma$ diagram. Using numerical models it is possible to reliably identify regions on the diagram that correspond to young bubbles, and more accurately determine their age and the properties of the surrounding gas. The region of the diagram with low H$\alpha$ brightness for both low and high values of velocity dispersion requires a more detailed analysis of the components since this region includes parts of supershells of different ages.

\section*{acknowledgements}
The authors are grateful to O.V.~Egorov and A.V.~Moiseev for numerous discussions and valuable clarifications.


\end{document}